\documentclass[10pt,a4paper]{article}

\usepackage{authblk}
\usepackage[utf8]{inputenc}
\usepackage[round]{natbib}
\usepackage[pdftex]{graphicx}
\usepackage{amsmath,amsthm,amsfonts}
\usepackage{lscape}   
\usepackage{rotating}
\usepackage{rotfloat}
\usepackage{multirow}
\usepackage{booktabs}
\usepackage{ctable}
\usepackage{threeparttable}
\usepackage{booktabs}
\usepackage{framed}
\usepackage{color}
\usepackage[colorlinks=true,linkcolor=blue,allcolors=blue]{hyperref}

\usepackage{soul}
\usepackage[ruled,vlined]{algorithm2e}
\usepackage{listings}
\usepackage{amsmath}
\usepackage{authblk}
\newtheorem{theorem}{Theorem}

\newtheorem{corollary}{Corollary}
\newtheorem{definition}{Definition}

\title{\texttt{FRCI}: An \texttt{R} Package for fuzzy and randomized confidence intervals}
\author{Carlos Henrique Trigo Nasser Felix, Nancy Lopes Garcia and Alex Rodrigo dos Santos Sousa \thanks{asousa@unicamp.br}}
\affil{Universidade Estadual de Campinas \\ Department of Statistics, Brazil}

\date{} 

\begin{document}

\maketitle

\vspace{-1.0cm}
\begin{abstract}

  Fuzzy confidence intervals provide a flexible framework for statistical inference by representing uncertainty through membership functions rather than fixed interval endpoints. Closely related to randomized confidence intervals, they offer an alternative approach for dealing with the conservativeness that frequently arises in discrete statistical models. Despite their attractive theoretical properties, the practical use of fuzzy and randomized confidence intervals has been limited by the lack of readily available software implementations.

This paper introduces \texttt{FRCI} (Fuzzy and Randomized Confidence Intervals), an \texttt{R} package available on CRAN for constructing and visualizing fuzzy and randomized confidence intervals. The package implements confidence interval procedures for parameters of the normal, Bernoulli/binomial, and Poisson distributions, including a recently proposed method based on reference value choice, Geyer--Meeden, Agresti--Coull, and score methods. The main function computes the membership function associated with a selected confidence interval procedure, while dedicated graphical tools facilitate visualization and comparison of fuzzy confidence intervals. We describe the statistical background underlying the implemented methods, discuss the package design, and illustrate its use through several examples. The \texttt{FRCI} package provides researchers, practitioners, and students with an accessible and unified environment for the study and application of fuzzy and randomized confidence intervals.
    \\

\noindent{\bf Keywords:} fuzzy confidence interval, randomized confidence interval, membership function, statistical inference, \texttt{R}.\\
  \end{abstract}

\section{Introduction} \label{sec:intro}

Confidence intervals are among the most widely used tools in statistical inference, providing a range of plausible values for an unknown parameter based on observed data. Classical confidence intervals are usually constructed to achieve a prescribed coverage probability while maintaining desirable properties such as short expected length and computational simplicity. However, in discrete sampling models, such as the binomial and Poisson distributions, exact confidence intervals often exhibit conservative behavior, leading to coverage probabilities substantially larger than the nominal confidence level. This phenomenon is a consequence of the discrete nature of the sample space and has motivated the development of alternative inferential procedures.

Randomized confidence intervals constitute one approach to reducing conservativeness in discrete problems. By incorporating an auxiliary random variable into the interval construction, these methods can achieve exact nominal coverage probabilities while often producing shorter intervals than their non-randomized counterparts. Despite their theoretical appeal, randomized intervals are not commonly used in practice because their interpretation is often considered unintuitive and their implementation may be cumbersome. See \cite{casella2024} for a general overview on confidence and randomized confidence intervals.

An alternative framework is provided by fuzzy confidence intervals, originally proposed to combine ideas from fuzzy set theory and statistical inference. Instead of representing uncertainty through a single interval, fuzzy confidence intervals associate each parameter value with a membership degree, producing a fuzzy set on the parameter space. This representation can be interpreted as a continuum of confidence intervals indexed by the confidence level, providing a richer description of inferential uncertainty. Furthermore, fuzzy confidence intervals preserve many desirable properties of randomized confidence intervals while avoiding the need to generate auxiliary random variables in practical applications. See \cite{fuzzy} for an overview on fuzzy logic.

Some fuzzy confidence interval constructions have been proposed in the literature for different statistical models and interval estimation procedures, see \cite{dollinger}, \cite{Geyer2005} and \cite{holladay2019}. Nevertheless, their practical use remains limited due to the lack of readily available software tools. Researchers and practitioners interested in applying these methods often need to implement the underlying algorithms from scratch, which may hinder their adoption and reproducibility.

To address this gap, this paper introduces  \texttt{FRCI} (Fuzzy and Randomized Confidence Intervals) package for \texttt{R} available on CRAN \citep{package}. The package provides computational tools for constructing and visualizing fuzzy and randomized confidence intervals for parameters of the normal, binomial, and Poisson distributions. The main function generates the membership function associated with a chosen confidence interval procedure, while dedicated graphical tools allow users to inspect and compare fuzzy confidence intervals under different sampling models and methods. The package aims to make fuzzy confidence interval methodology easily accessible to statisticians, researchers, and students through a simple and unified interface.

The remainder of this paper is organized as follows. Section~\ref{sec:theory} briefly reviews the theoretical foundations of fuzzy and randomized confidence intervals. Section~\ref{sec:functions} describes the main functions implemented in \texttt{FRCI}. Illustrative examples are presented in Section~\ref{sec:examples}, and concluding remarks are given in the final section.



\section{Theoretical background} \label{sec:theory}
In this section, we provide a brief background on fuzzy confidence intervals and describe the methods implemented in the \texttt{FRCI} package. In particular, the first two methods are fuzzy: the first was proposed by \cite{carlos2025}, and the second by \cite{Geyer2005}. The last two methods are approximate procedures for constructing confidence intervals: one for the proportion parameter of the Binomial distribution, proposed by \cite{Agresti1998}, and another for the mean of the Poisson distribution based on the score method.

\subsection{Fuzzy confidence intervals}
A fuzzy set $A$ in a space $S$ is characterized by its membership function $\psi: S \rightarrow [0,1]$, where $\psi(x)$ represents how much we believe $x \in A$. When $\psi(x)$ is closer to $1$, the more we believe $x \in A$ and, in the opposite, when $\psi(x)$ is closer to $0$, less we think $x \in A$. The $\gamma$-cut of the fuzzy set $A$, denoted by $\Psi_{A}^{\gamma}$, is
\begin{equation} \label{cut}
    \Psi_A^\gamma = \{x \in S: \psi(x)\geq\gamma\}, 
\end{equation}
$\gamma \in [0,1]$.

Let $\Theta$ denote a parametric space, and consider $(\Omega,\mathcal A,\mu(\cdot \mid \theta))$ a probability space indexed by $\theta \in \Theta$ and denote $\mathbb{E}_{\theta}$ the corresponding expectation functional. Let $(\Theta, \mathcal{O}, \nu)$ be a measure space defining the size on $\Theta$. Let $\gamma \in (0,1)$ be the desired confidence level.

\begin{definition} [Fuzzy confidence interval and expected length]\label{def:1} An $\mathcal{A}$-measurable membership function $\psi : \Omega \times \Theta \to [0,1]$ satisfying
\[
\int_{\Omega} \psi(\omega \mid \tau)\, d\mu(\omega \mid \tau) \ge \gamma, \, \mbox{ for all }  \, \tau \in \Theta, 
\]
defines a fuzzy confidence interval with $100\gamma\%$ confidence.  The randomized rejection region for the Neyman-Pearson test for $H_0: \theta=\tau$ with significance level $1-\gamma$ is given by the rejection function $\psi(\tau, \cdot): \Omega \rightarrow [0,1]$. The quantity
\[
\mathrm{EL}(\psi,\theta, \nu)
= \mathbb{E}_{\theta}\!\left[\int_{\Theta} \psi(\omega \mid \tau)\, d\nu(\tau)\right]
\]
is called the expected length of the fuzzy confidence interval.
\end{definition}

See \cite{dollinger} and \cite{holladay2019} for a complete overview about fuzzy confidence intervals. The next subsections describe the methods provided by the proposed R package \texttt{FRCI} for constructing fuzzy and approximate confidence intervals for the proportion parameter of the Binomial distribution and the mean of the Poisson distribution and fuzzy confidence interval for the bounded mean of the normal distribution.

\subsection{Proposed method}

The approach consists of selecting a reference value $o \in \Theta$ and, given this choice, constructing a function 
$\psi_o:\Omega\times\Theta \to [0,1]$
that minimizes the probability of failing to reject $H_0: \theta= \tau$ when the distribution associated with $o$ is taken as the true one, for any $\tau \neq o$. For $\psi_o$ to define a confidence interval, as required in Definition \ref{def:1}, the probability of failing to reject $H_0$ when the distribution associated with $\tau$ is the true one be greater than or equal to $\gamma$, for all $\tau \in \Theta$. This condition serves as the constraint of the minimization problem. The constraint can be expressed directly on the set over which the functions 
$\psi$ are evaluated. Specifically, we aim to minimize 
\begin{equation*}
   \int_\Omega \psi(\omega \mid \tau)\, d\mu(\omega \mid o), \, \mbox{ for } \, \psi \in \mathfrak{F}_\gamma,
\end{equation*}
 where
\begin{eqnarray*}
\mathfrak{F_\gamma} &=& \left\{\psi:\Omega\times\Theta\to [0,1]\ \Bigg|\  \psi(\cdot \mid \tau)\ \text{is}\ \mathcal{A}/\mathcal{B}(\mathbb{R})\text{-measurable and } \right. \\
&&\hspace{5cm} \left.\int_\Omega\psi(\omega \mid \tau)d\mu(\omega \mid \tau)\ge \gamma , \forall\tau\in\Theta \right\}.
\end{eqnarray*}

First, we will present the nonparametric version of the theorem of interest. The parametric version follows as a corollary.

\begin{theorem}\label{thm:1} Let $(\Omega,\mathcal{A},\mu)$ and $(\Omega,\mathcal{A},\nu)$ be two probability spaces, and let $\gamma\in(0,1)$ denote the desired confidence level. Define
\begin{eqnarray*}
\mathfrak{E}(\gamma,\mu) &=& \left\{\psi:\Omega\to [0,1]\ \Bigg|\  \psi\ \text{is}\ \mathcal{A}\text{-measurable and } \int_\Omega \psi(\omega)\, d\mu(\omega)\ge \gamma \right\}
\end{eqnarray*}
as the set of $\mathcal{A}$-measurable functions with coverage probability greater than $\gamma$. Then there exists $\psi^*\in\mathfrak{E}(\gamma,\mu)$ such that, for every $\psi\in\mathfrak{E}(\gamma,\mu)$,
$$
\int_\Omega \psi(\omega)\, d\nu(\omega)
\ge
\int_\Omega \psi^*(\omega)\, d\nu(\omega).
$$
\end{theorem}

To construct the randomized confidence interval, we use some interpretations of Theorem \ref{thm:1}, which can be seen as a hypothesis test with simple hypotheses, with $H_0: \mu$ describing the distribution of the data versus $H_1: \nu$ describing the distribution of the data. We can parametrize these measures in a parametric family while maintaining the property of being the Uniformly Most Powerful (UMP) test.

Consider the parametric space $\Theta$ and the values $o$ and $\tau$ $\in \Theta$, writing the associated probability measures as $\mu(\cdot \mid o)$ and $\mu(\cdot \mid \tau)$, respectively. We can obtain the membership function given $o$ by combining the functions indexed by $\tau$ defined as the minimizing function $\psi_\tau^*:\Omega\to[0,1]$ with $\psi_\tau^*\in \mathfrak{E}(\gamma,\mu(\cdot \mid \tau))$, $\mu=\mu(\cdot \mid \tau)$, and $\nu=\mu(\cdot \mid o)$, defining $\psi_o(\omega \mid \tau)=\psi_\tau^*(\omega)$. Therefore, $\psi_o\in\mathfrak{F_\gamma}$ and $\psi_o$ is UMP for $H_0: \theta=o$.

\begin{corollary}\label{thm:2} Let $\Theta$ be a parametric space, $(\Omega,\mathcal{A},\mu(\cdot \mid \tau))$ a probability space for all $\tau\in\Theta$, and $\gamma\in(0,1)$ the desired confidence level. Let 
\begin{eqnarray*}
\mathfrak{F_\gamma} &=& \left\{\psi:\Omega\times\Theta\to [0,1]\ \Bigg|\  \psi(\cdot \mid \tau)\ \text{is}\ \mathcal{A}/\mathcal{B}(\mathbb{R})\text{-measurable and } \right. \\
&&\hspace{5cm} \left.\int_\Omega\psi(\omega \mid \tau)d\mu(\omega|\tau)\ge \gamma , \forall\tau\in\Theta \right\}
\end{eqnarray*}
be the set of $\mathcal{A}$-measurable functions with coverage rate at least $\gamma$. Then, for a given $o\in\Theta$, there exists $\psi^*\in\mathfrak{F_\gamma}$ such that for all $\psi\in\mathfrak{F_\gamma}$,
$$\int_\Omega\psi(\omega \mid \tau)d\mu(\omega \mid o)\ge\int_\Omega\psi^* (\tau|\omega)d\mu(\omega \mid o).$$
\end{corollary}

This theorem requires fewer conditions than the Neyman-Pearson Lemma, but in exchange, the uniqueness property is lost. 


\begin{theorem}\label{thm:3} Given the conditions and definitions in Theorem \ref{thm:1}, consider the measure space $(\Theta,\mathcal{O},\nu)$ such that, for every $o\in\Theta$,
$$(\theta,\tau)\mapsto\int_\Omega\psi_{o}(\omega \mid \tau)d\mu(\omega\mid\theta)$$
is $\mathcal{O}$-measurable. Let 
the subset $\mathfrak{M_\gamma}$ of $\mathfrak{F_\gamma}$ be defined by 
\begin{equation*}
    \mathfrak{M_\gamma}=\left\{\psi\in\mathfrak{F_\gamma}\ \middle|\ (\theta,\tau)\mapsto\int_\Omega\psi(\omega \mid \tau)d\mu(\omega\mid\theta)\mbox{ is $\mathcal{O}$-measurable}\right\}
\end{equation*}

and define the function $\mathrm{EL}(\theta,\psi):\Theta\times\mathfrak{M_\gamma}\to[0,+\infty]$ by 
$$\mathrm{EL}(\theta,\psi)=\int_\Theta\int_\Omega\psi(\omega \mid \tau)d\mu(\omega\mid\theta)d\nu(\tau),$$
then for every $\gamma\in(0,1)$ and $\theta\in\Theta$ there exists $\psi^*\in\mathfrak{M_\gamma}$ such that
\begin{equation*}
    \mathrm{EL}(\theta,\psi)\ge \mathrm{EL}(\theta,\psi^*), \mbox{ for all $\psi\in\mathfrak{M_\gamma}$.}
\end{equation*}

\end{theorem}

The proofs of the Theorems \ref{thm:1} and \ref{thm:3} and Corollary \ref{thm:2} are provided in \cite{carlos2025}. In the \texttt{FRCI} package, the proposed procedure is implemented to construct fuzzy confidence intervals for the proportion parameter of the Binomial distribution, the mean of the Poisson distribution and the mean of the Normal distribution.

\subsection{Geyer-Meeden fuzzy confidence interval}
Let $\phi(x,\alpha,\theta)$ be the critical function of a randomized test, where $x$ is the data, $\alpha$ is the significance level and $\theta$ is the hypothesized value of the parameter under the null hypothesis. For instance, in a UMP test $H_0: \theta \leq \theta_0$ versus $H_1: \theta > \theta_0$ for a one-parameter model with monotone likelihood ratio function in $T(x)$,
\begin{equation}\label{ump}
\phi(x,\alpha,\theta_0) =
\begin{cases}
1, & T(x)>C \\
\gamma, & T(x)=C \\
0, & T(x)<C,
\end{cases}
\end{equation}
where $\gamma$ and $C$ are such that $\mathbb{E}_{\theta_0}[\phi(X,\alpha,\theta)] = \alpha$. Thus, the fuzzy confidence interval is defined by the membership function 
\begin{equation} \label{geyer}
    \psi(\theta) = 1 -  \phi(x,\alpha,\theta),
\end{equation} 
for $x$ and $\alpha$ fixed. \cite{Geyer2005} propose for statisticians and scientists in general to report the value of \eqref{geyer} instead of conducting the randomization procedure. Further, they implemented their proposal for constructing a fuzzy confidence interval for the proportion parameter of the Binomial distribution in the R package \texttt{ump} (\cite{gm_pack}). 

\subsection{Agresti-Coull confidence interval for the proportion of the Binomial distribution} 

Let $X$ be a random variable with Binomial distribution with parameters $n \in \mathbb{N}$ and $p \in (0,1)$. The score confidence interval for $p$ with level $(1-\alpha)100\%$ is given by
\begin{equation} \label{score_bin}
\frac{\hat{p} + \frac{z_{\alpha/2}^2}{2n} \pm z_{\alpha/2}\sqrt{\frac{\hat{p}(1-\hat{p})}{n} + \frac{z_{\alpha/2}^2}{4n^2}}}{1 + \frac{z_{\alpha/2}^2}{n}},
\end{equation}
where $\hat{p} = X/n$ is the sample proportion and $z_{\alpha/2}$ is the $(1 - \alpha/2)$ quantile of the standard normal distribution. The score confidence interval is constructed by inverting the score test, in which the standard error is evaluated using the log-likelihood at the null hypothesis value of $p$. In contrast, the Wald test uses a standard error based on the log-likelihood evaluated at the maximum likelihood estimate of $p$. \cite{Agresti1998} showed that the score confidence interval for $p$ tends to perform better than both the Wald confidence interval (obtained by inverting the Wald test) and the exact confidence interval in terms of coverage probability, as its coverage is typically closer to the nominal confidence level than that of the other two procedures. Further, they proposed an adaptation of the Wald confidence interval, called Agresti-Coull interval in \cite{brown2001}. Denote $\tilde{X} = X + z_{\alpha/2}^2/2$, $\tilde{n} = n + z_{\alpha/2}^2$ and $\tilde{p} = \tilde{X}/\tilde{n}$, then the Agresti-Coull $(1-\alpha)100\%$ confidence interval for $p$ is
\begin{equation} \label{ac}
    \tilde{p} \pm z_{\alpha/2}\sqrt{\frac{\tilde{p}(1-\tilde{p})}{\tilde{n}}}.
\end{equation}
See \cite{brown2001} and \cite{brown2002} for overviews about confidence intervals for the parameter $p$ of the Binomial distribution. Further, the Agresti-Coull procedure \eqref{ac} is also implemented in some R packages, such as \texttt{binom} (\cite{ac_pack1}) and \texttt{DescTools} (\cite{ac_pack2}.

\subsection{Score confidence interval for the mean of the Poisson distribution}

Let $X_1,\cdots,X_n$ be independent and identically distributed (IID) random variables with Poisson distribution with parameter $\lambda > 0$. The score confidence interval for $\lambda$ with level $(1-\alpha)100\%$ is 
\begin{equation}\label{score_pois}
    \bar{X} + \frac{z_{\alpha/2}^2}{2n} \pm z_{\alpha/2}\sqrt{\frac{1}{4n^2}(z_{\alpha/2}^2 + 4n\bar{X})},
\end{equation}
where $\bar{X} = (1/n)\sum_{i=1}^n X_i$ is the sample mean and $z_{\alpha/2}$ is the $(1-\alpha/2)$ quantile of the standard normal distribution, see \cite{nadarajah2015}. See also \cite{byrne2005} and \cite{swift2009} for comparisons of methods to obtain confidence intervals for the parameter $\lambda$ of the Poisson distribution and \cite{casella2024} for an overview about the score method in general. Further, the score confidence interval \eqref{score_pois} is implemented in R packages such as \texttt{DescTools} (\cite{ac_pack2} and \texttt{ratesci} (\cite{scorpoi_pack}.

The score confidence interval for the mean parameter of the Poisson distribution offers a superior balance between accuracy and efficiency when compared to classical alternatives such as the Wald and exact intervals. Unlike the Wald interval, which relies on a plug-in variance estimate and often performs poorly in small samples or low-count regimes, sometimes even producing negative lower bounds, the score interval is derived from the inversion of the score test and accounts more appropriately for the curvature of the likelihood function. As a result, it achieves coverage probabilities much closer to the nominal level. Additionally, while exact intervals guarantee coverage, they tend to be overly conservative and unnecessarily wide. In contrast, the score interval typically yields shorter intervals without sacrificing reliability, making it a robust and practically appealing choice for inference in Poisson models.

\section{Functions}
\label{sec:functions}

The \texttt{FRCI} package provides two main functions: \texttt{psi()}, which constructs membership functions associated with fuzzy confidence intervals, and \texttt{plot.psi()}, an S3 method for visualization and comparison of the resulting fuzzy confidence intervals.

\subsection{psi(): Construction of Membership Functions}

The function \texttt{psi()} is the core routine of the package. It generates the membership function corresponding to a fuzzy confidence interval for a parameter of interest under a specified sampling distribution and confidence interval construction method. The available arguments are summarized in Table \ref{tab:psi_args}.

The argument distribution specifies the sampling model. Currently, three distributions are supported: normal (\texttt{"normal"}), Bernoulli/binomial (\texttt{"bernoulli"}), and Poisson (\texttt{"poisson"}).

The argument \texttt{method} determines the confidence interval procedure used in the construction of the membership function. This argument can either be a numeric value or the name of a predefined method. A numeric value is interpreted as a reference parameter in the fuzzy confidence interval construction. Alternatively, the user may select one of the implemented methods: \texttt{"GM"} for the Geyer--Meeden interval, \texttt{"AC"} for the Agresti--Coull interval (available only for binomial data), and \texttt{"SC"} for the score interval (available only for Poisson data).

The argument \texttt{n} corresponds to the sample size. For the normal model, the standard deviation must also be supplied through the argument \texttt{sigma}.

Typical uses of the function are illustrated below:

\begin{verbatim}
R> p_gm_binom <- psi(distribution = "bernoulli", method = "GM", n = 10)
R> p_0_norm <- psi(distribution = "normal", method = 0, n = 1, sigma = 10)
\end{verbatim}

The output is an object of class \texttt{"psi"}, which contains all information required to evaluate and visualize the membership function. In particular, the object stores the membership function itself, the selected sampling distribution, the sample size, the confidence interval method, and the support of the membership function. Objects of class \texttt{"psi"} can be directly supplied to the plotting method \texttt{plot.psi()}.

\begin{table}[ht]
\centering
\caption{Arguments and their descriptions of the \texttt{psi()} function.}
\label{tab:psi_args}
\begin{tabular}{p{2.2cm}p{11cm}}
\hline
Argument & Description \\
\hline
\texttt{distribution} &
Sampling distribution. Available options are
\texttt{"normal"},
\texttt{"bernoulli"},
and
\texttt{"poisson"}. \\

\texttt{method} &
Confidence interval procedure. Available options are a numeric reference value,
\texttt{"GM"} (Geyer--Meeden),
\texttt{"AC"} (Agresti--Coull, Bernoulli only),
and
\texttt{"SC"} (score interval, Poisson only). \\

\texttt{n} &
Sample size. \\

\texttt{sigma} &
Known standard deviation, required only for the normal distribution. \\
\hline
\end{tabular}
\end{table}

\subsection{plot.psi(): Visualization of Fuzzy Confidence Intervals}

The function \texttt{plot.psi()} is an S3 plotting method for objects of class \texttt{"psi"}. It provides graphical representations of the membership function and related quantities associated with fuzzy confidence intervals. The arguments and their descriptions are provided in Table \ref{tab:plotpsi_args}.

\begin{table}[ht]
\centering
\caption{Arguments and their descriptions of the \texttt{plot.psi()} function.}
\label{tab:plotpsi_args}
\begin{tabular}{p{2.2cm}p{11cm}}
\hline
Argument &  Description \\
\hline
\texttt{x} & A list that contains: membership function, sample distribution, sample size, method, standard deviation (for normal case), upper and lower bound of the support of the membership function. \\ 

\texttt{...} & More lists that contains: membership function, sample distribution, sample size, method, standard deviation (for normal case), upper and lower bound of the support of the membership function. \\ 

\texttt{gamma} & The confidence level. \\

\texttt{type} & The type of the plot, by default is "psi" for the membership function, and "length" for the expected length plot. \\ 

\texttt{xlim} & A vector containing the upper and lower values for the x-axis of the plot. \\

\texttt{tlim} & A vector containing the upper and lower values for $\theta$\\

\texttt{yval} & A vector containing the values of omega to be included in the plot. (for type = "psi" only)\\

\texttt{ylim} & A vector containing the upper and lower values for the y-axis of the plot. (for type = "length" only)\\

\hline

\hline
\end{tabular}
\end{table}

The main argument, \texttt{x}, is a \texttt{"psi"} object generated by \texttt{psi()}. An optional second object, supplied through \texttt{psi2}, allows the comparison of two fuzzy confidence intervals within the same graph. The confidence level used in the visualization is specified through \texttt{gamma}.

Additional graphical control is available through the arguments \texttt{xlim} and \texttt{ylim}, which define the ranges of the horizontal and vertical axes, respectively. The argument \texttt{type} determines the type of graphical output. By default, the membership function is displayed, whereas setting \texttt{type = "length"} produces a plot of the expected interval length as a function of the confidence level.

\begin{figure}[h]
\centering
\includegraphics[scale=0.6]{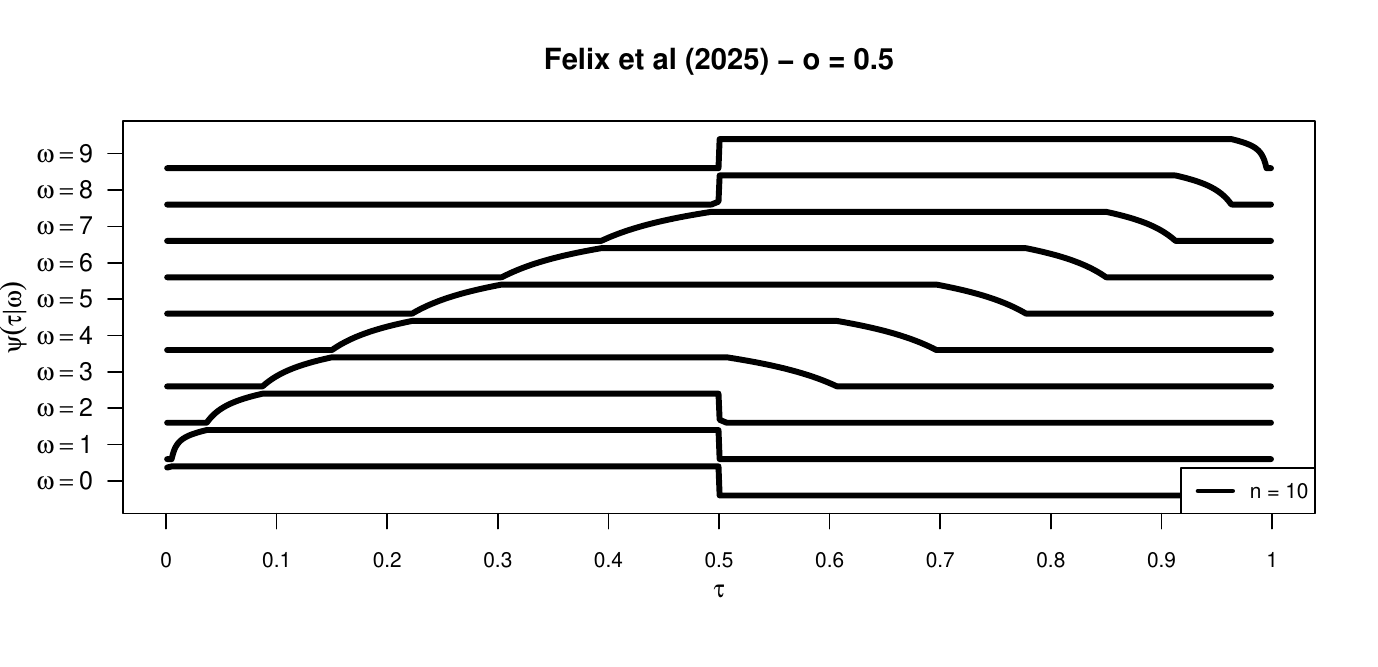}
\caption{Membership function of the proposed method for binomial distribution with $n = 10$ and $o = 0.5$.}
\label{fig:bin10ex}
\end{figure}

The following example generates and displays the membership function associated with the proposed fuzzy confidence interval for a Binomial proportion with $n = 10$ and $o = 0.5$:
\begin{verbatim}
R> psi_gm_binom <- psi(distribution = "bernoulli", method = 0.5, n = 10)
R> plot(psi_gm_binom)
\end{verbatim}

The output is showed in Figure \ref{fig:bin10ex}. Such visualizations provide an intuitive representation of the uncertainty associated with the parameter of interest and facilitate comparisons among different confidence interval procedures. 

\section{Examples} 
\label{sec:examples}

In this section, we present several examples illustrating the use of the package to obtain the confidence intervals and their expected lengths for binomial proportions, Poisson and normal means. 

\subsection{Binomial proportion} 
Let us consider the membership functions for the binomial proportion for sample sizes $n=1,2,4,$ and $8$. For the proposed method, we consider $o=0.5$ and $o=0.75$. The corresponding membership functions are obtained using the \texttt{psi} function by setting the argument \texttt{method = .5} and \texttt{method = .75}, respectively. In both cases, the argument \texttt{distribution = "bernoulli"} is used to specify the binomial distribution, as follows:

\begin{verbatim}
R> psi_binom_o_.5_1<-psi(distribution = "bernoulli",method = .5,n = 1)
R> psi_binom_o_.5_2<-psi(distribution = "bernoulli",method = .5,n = 2)
R> psi_binom_o_.5_4<-psi(distribution = "bernoulli",method = .5,n = 4)
R> psi_binom_o_.5_8<-psi(distribution = "bernoulli",method = .5,n = 8)

R> psi_binom_o_.75_1<-psi(distribution = "bernoulli",method = .75,n = 1)
R> psi_binom_o_.75_2<-psi(distribution = "bernoulli",method = .75,n = 2)
R> psi_binom_o_.75_4<-psi(distribution = "bernoulli",method = .75,n = 4)
R> psi_binom_o_.75_8<-psi(distribution = "bernoulli",method = .75,n = 8)
\end{verbatim}

\begin{figure}[h] 
\includegraphics[width=0.24\textwidth]{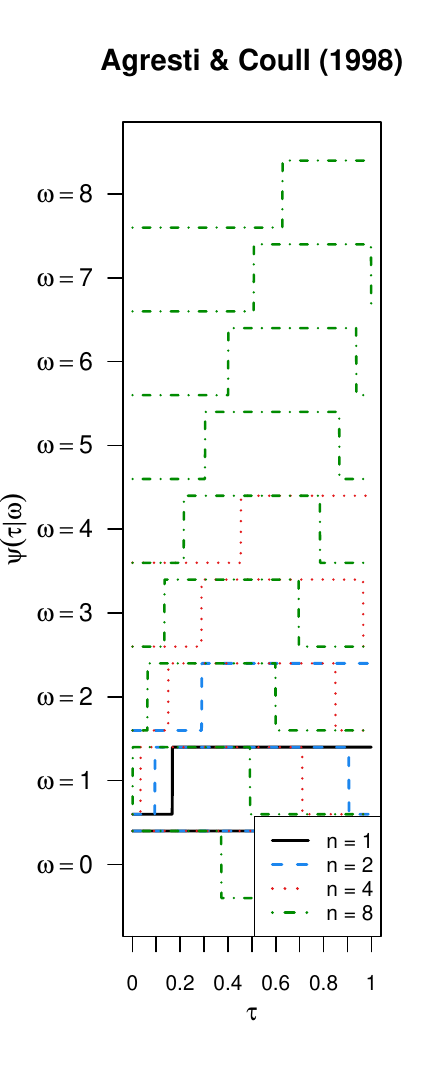}
\includegraphics[width=0.24\textwidth]{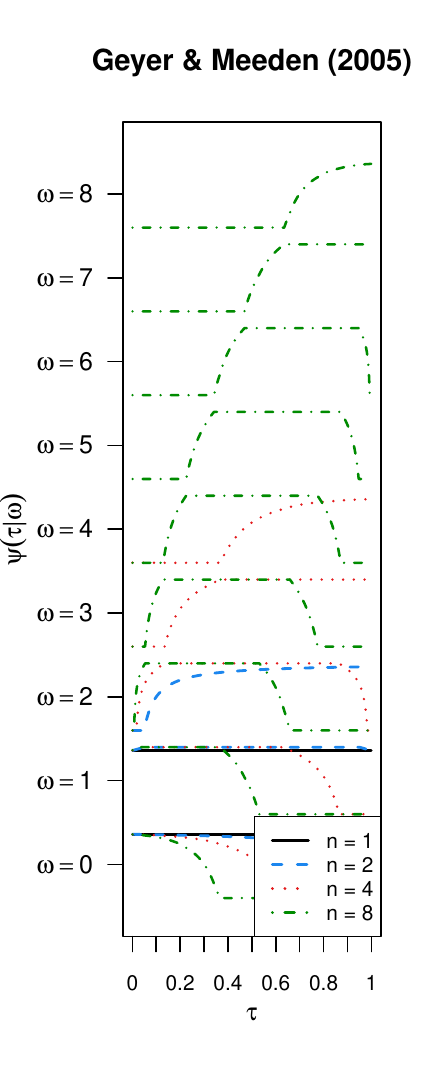}
\includegraphics[width=0.24\textwidth]{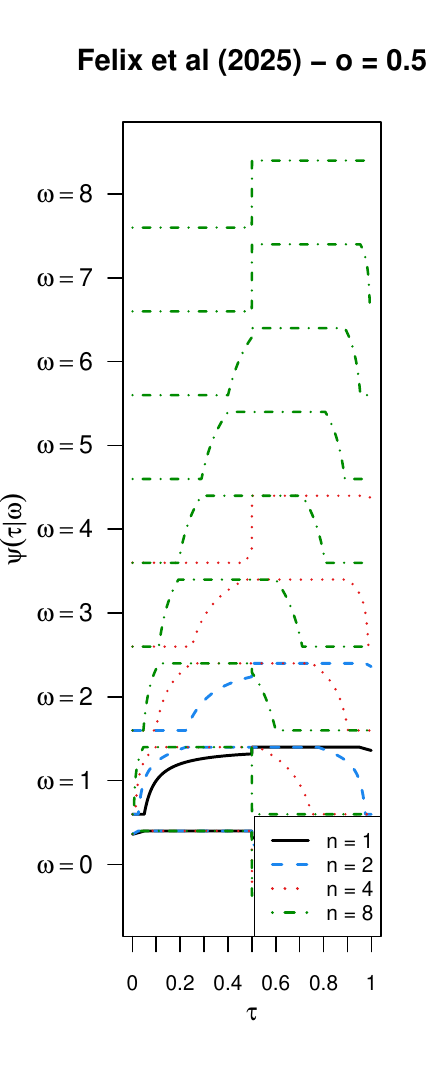}
\includegraphics[width=0.24\textwidth]{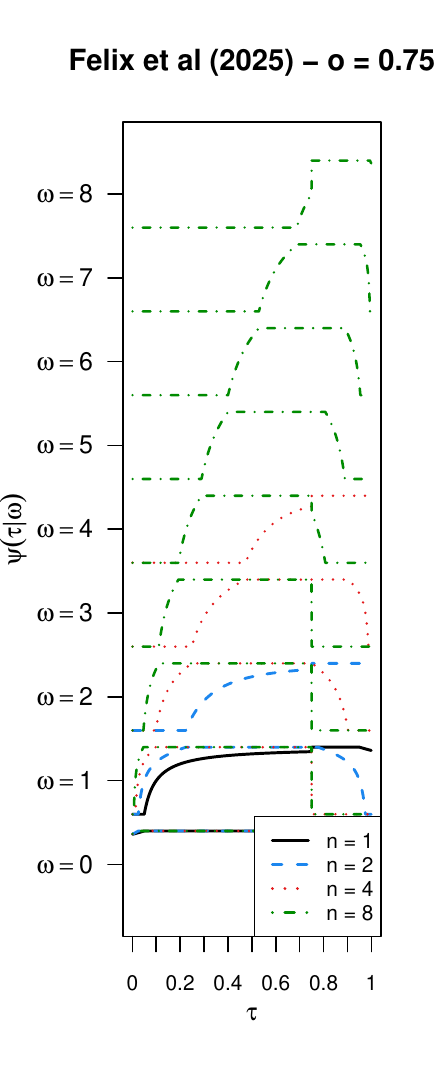}
\caption{Membership functions for the binomial proportion obtained using the proposed method with $o = 0.5$ and $o = 0.75$, as well as the Agresti--Coull and Geyer--Meeden methods. The plots were generated using the \texttt{plot.psi} function with a confidence level of $\gamma = 0.95$ and sample sizes $n = 1, 2, 4,$ and $8$.
}\label{memplot_bin}
\end{figure}

For the Geyer--Meeden and Agresti--Coull methods, we change the \texttt{method} argument in the \texttt{psi} function by setting \texttt{method = "GM"} and \texttt{method = "AC"}, respectively,

\begin{verbatim}
R> psi_binom_gm_1<-psi(distribution = "bernoulli",method = "GM",n = 1)
R> psi_binom_gm_2<-psi(distribution = "bernoulli",method = "GM",n = 2)
R> psi_binom_gm_4<-psi(distribution = "bernoulli",method = "GM",n = 4)
R> psi_binom_gm_8<-psi(distribution = "bernoulli",method = "GM",n = 8)

R> psi_binom_ac_1<-psi(distribution = "bernoulli",method = "AC",n = 1)
R> psi_binom_ac_2<-psi(distribution = "bernoulli",method = "AC",n = 2)
R> psi_binom_ac_4<-psi(distribution = "bernoulli",method = "AC",n = 4)
R> psi_binom_ac_8<-psi(distribution = "bernoulli",method = "AC",n = 8)
\end{verbatim}
We now use the \texttt{plot} function to provide graphical representations of the membership functions for $\omega=0,\ldots,8$, specified through the argument \texttt{yval = 0:8} and the default confidence level $\gamma = 0.95$ as follows:
\begin{verbatim}
R> plot(psi_binom_ac_1,psi_binom_ac_2,psi_binom_ac_4,psi_binom_ac_8,
     psi_binom_gm_1,psi_binom_gm_2,psi_binom_gm_4,psi_binom_gm_8,
     psi_binom_o_.5_1,psi_binom_o_.5_2,psi_binom_o_.5_4,psi_binom_o_.5_8,
     psi_binom_o_.75_1,psi_binom_o_.75_2,psi_binom_o_.75_4,psi_binom_o_.75_8,
     yval = 0:8)   
\end{verbatim}
The resulting plots are shown in Figure~\ref{memplot_bin}. As $n$ increases, the membership functions become more localized, 
reflecting the increasing amount of information available about the 
binomial proportion. Moreover, the regions of higher membership shift 
toward larger values of $\tau$ as the observed proportion $\omega/n$ 
increases. For small sample sizes, particularly $n=1$ and $n=2$, the 
membership functions are relatively broad, indicating greater uncertainty 
about the parameter. In contrast, for $n=4$ and especially for $n=8$, 
the functions exhibit more sharply defined regions, showing the increased 
precision resulting from larger sample sizes. For the proposed method, it is also possible to 
observe changes in the shape of the membership functions around 
$\tau=o$, which reflect the role of the reference value $o$ in the 
construction of the proposed fuzzy confidence sets. Overall, the figure 
shows that the membership functions appropriately respond to both the 
observed number of successes and the amount of information provided by 
the sample.



\begin{figure}[h]
\includegraphics[width=0.24\textwidth]{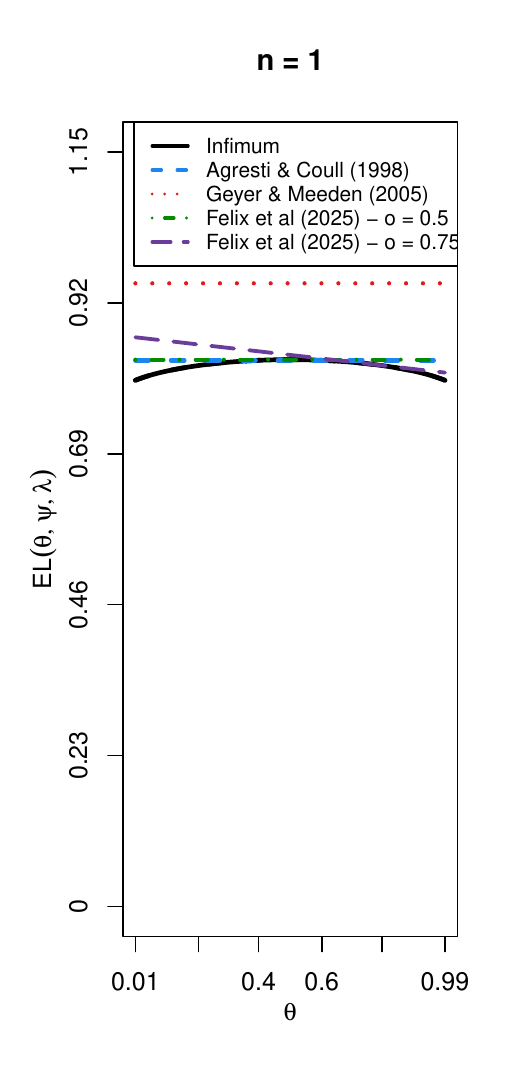}
\includegraphics[width=0.24\textwidth]{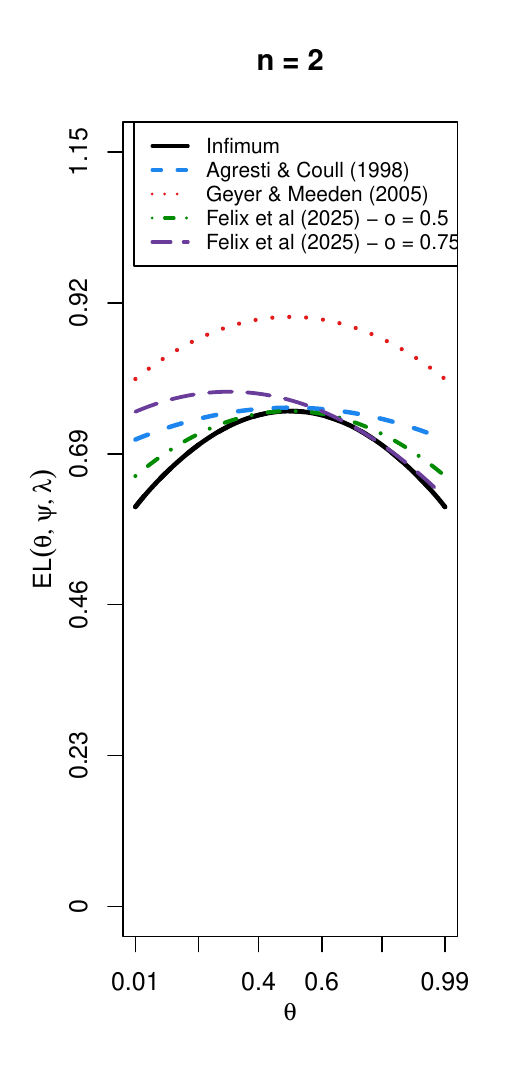}
\includegraphics[width=0.24\textwidth]{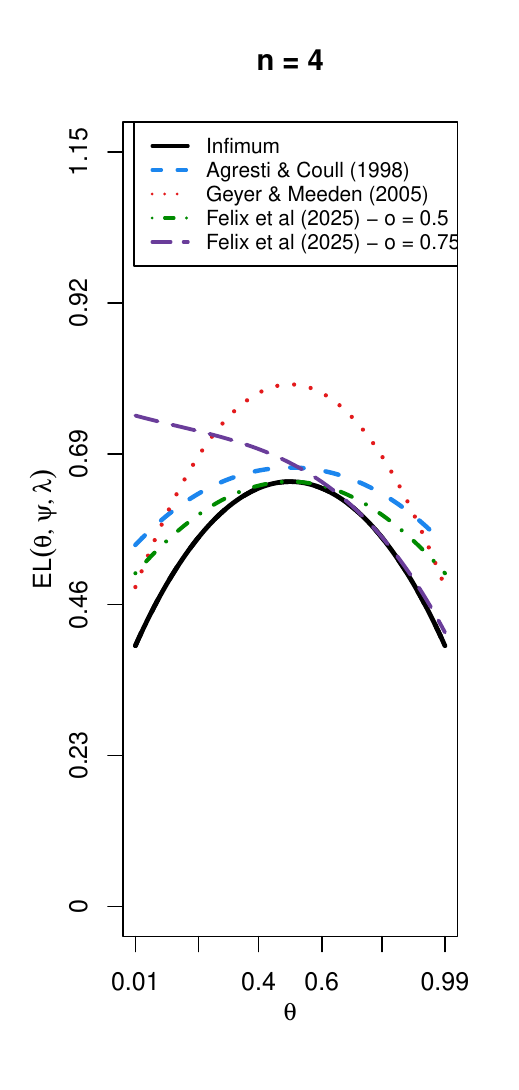}
\includegraphics[width=0.24\textwidth]{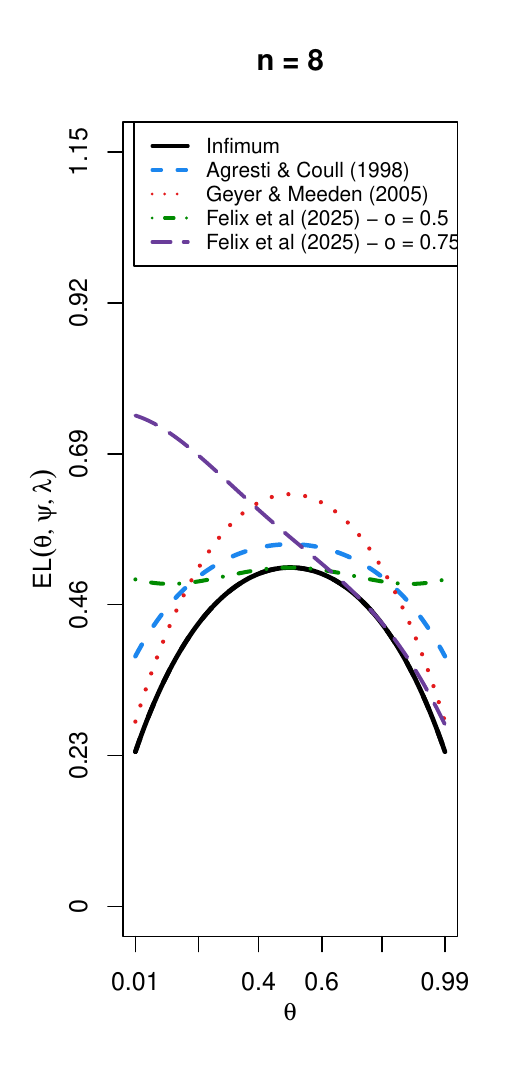}
\caption{Expected lengths of the interval for the binomial proportion using the proposed method with $o = 0.5$ and $o = 0.75$, as well as the Agresti--Coull and Geyer--Meeden methods. The plots were generated using the \texttt{plot.psi} function with a confidence level of $\gamma = 0.95$ and sample sizes $n = 1, 2, 4,$ and $8$.}\label{lengthplot_bin}
\end{figure}

The lengths are obtained specifying \texttt{type ="length"} in the \texttt{plot} function,
\begin{verbatim}
R> plot(psi_binom_ac_1,psi_binom_ac_2,psi_binom_ac_4,psi_binom_ac_8,
     psi_binom_gm_1,psi_binom_gm_2,psi_binom_gm_4,psi_binom_gm_8,
     psi_binom_o_.5_1,psi_binom_o_.5_2,psi_binom_o_.5_4,psi_binom_o_.5_8,
     psi_binom_o_.75_1,psi_binom_o_.75_2,psi_binom_o_.75_4,psi_binom_o_.75_8,
     type = "length",ylim=c(0,1.15))    
\end{verbatim}
and the plots are shown in Figure \ref{lengthplot_bin}. As expected from the construction of the proposed method, its expected
length coincides with the infimum at the corresponding reference value
$\theta=o$. Thus, the curve obtained with $o=0.5$ is tangent to the
infimum at $\theta=0.5$, whereas the curve corresponding to $o=0.75$
attains the infimum at $\theta=0.75$. Moreover, the proposed method
provides expected lengths close to the infimum in a neighborhood of the
chosen reference value.

%


\subsection{Poisson mean}

\begin{figure}[h]
\includegraphics[width=0.24\textwidth]{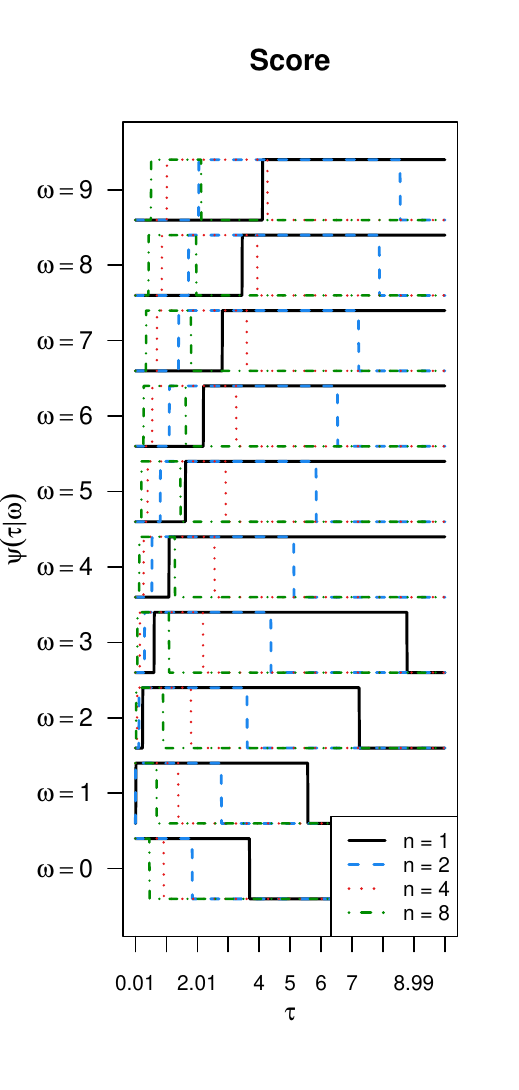}
\includegraphics[width=0.24\textwidth]{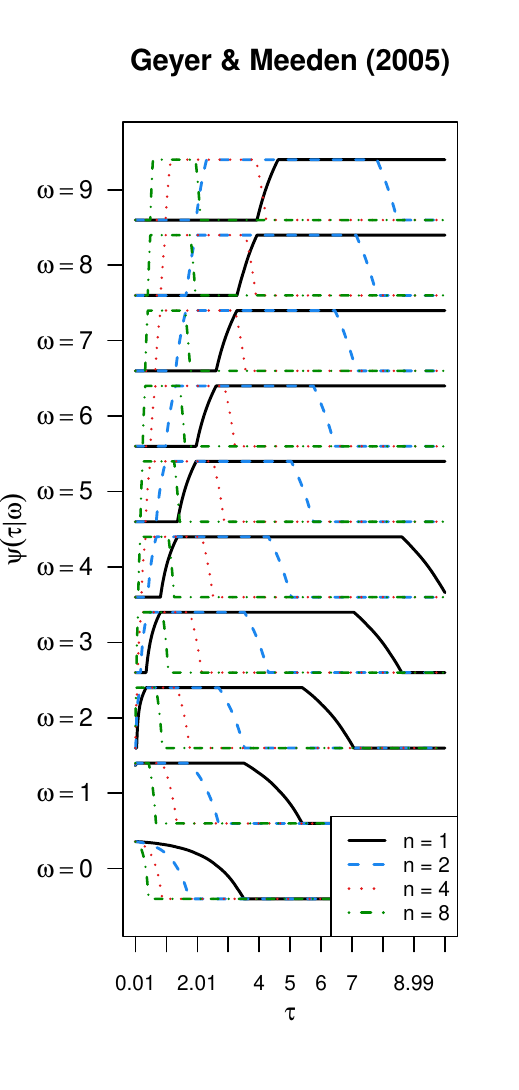}
\includegraphics[width=0.24\textwidth]{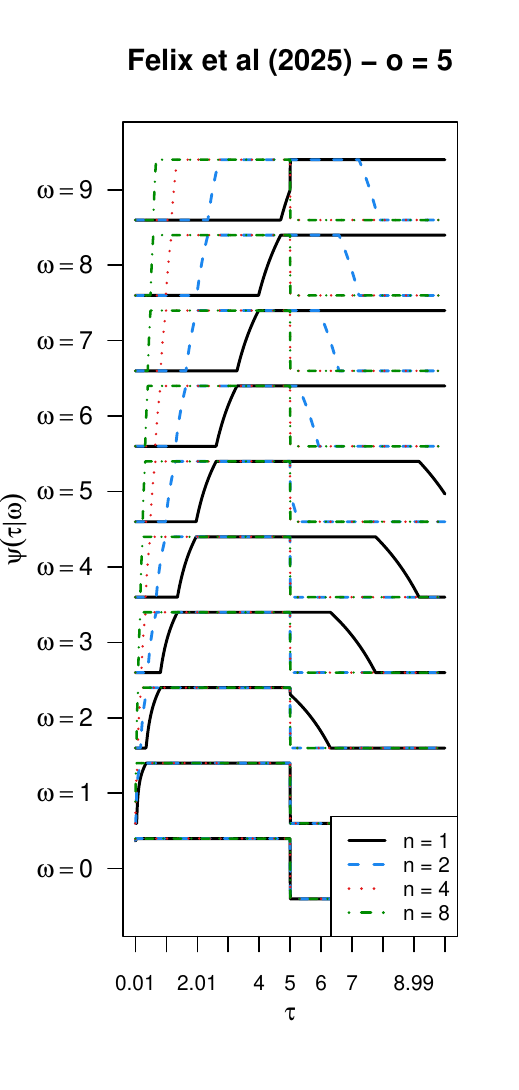}
\includegraphics[width=0.24\textwidth]{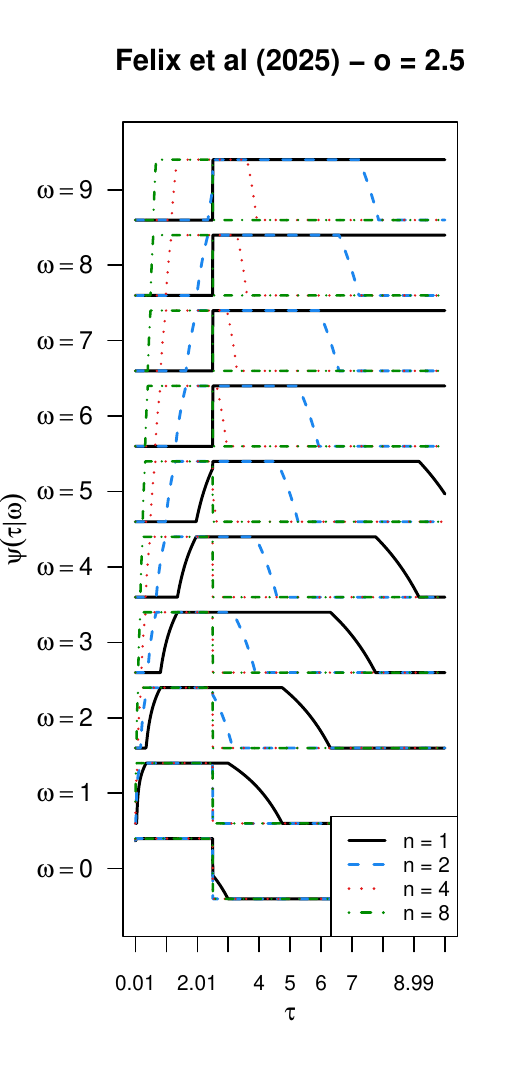}
\caption{Membership functions for the Poisson mean obtained using the proposed method with $o = 5$ and $o = 2.5$, as well as the score and Geyer--Meeden methods. The plots were generated using the \texttt{plot.psi} function with a confidence level of $\gamma = 0.95$ and sample sizes $n = 1, 2, 4,$ and $8$.
}\label{mempois_plot}
\end{figure}

\begin{figure}[h]
\includegraphics[width=0.24\textwidth]{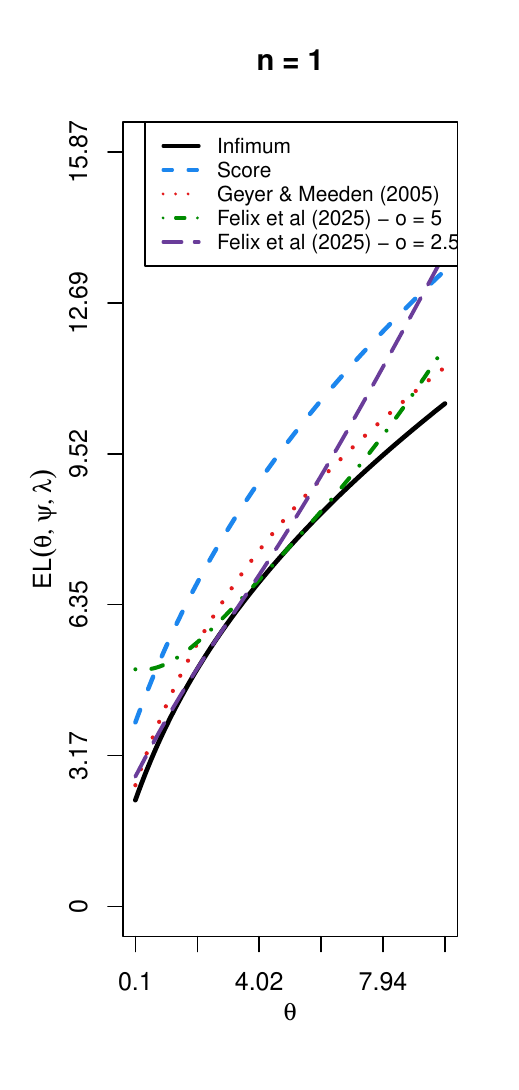}
\includegraphics[width=0.24\textwidth]{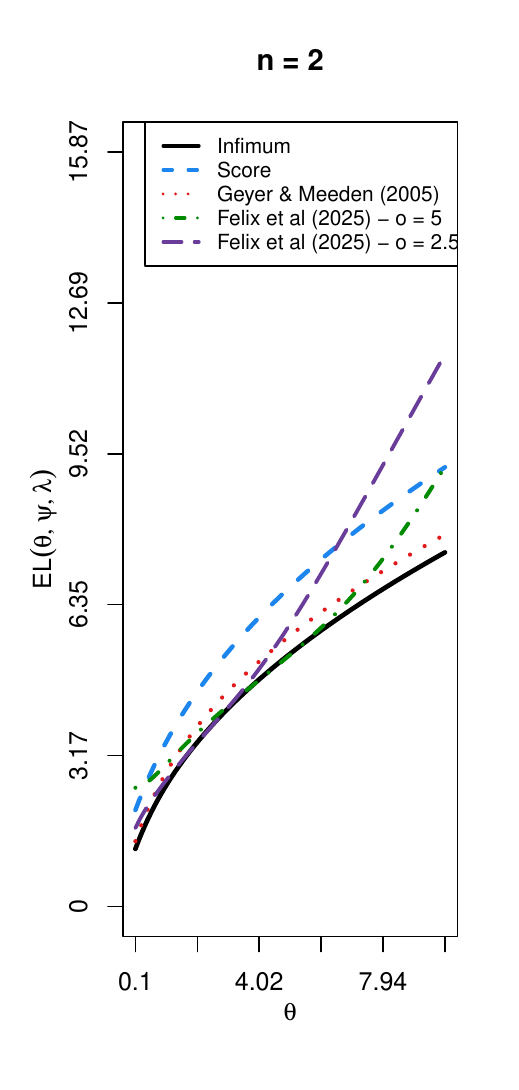}
\includegraphics[width=0.24\textwidth]{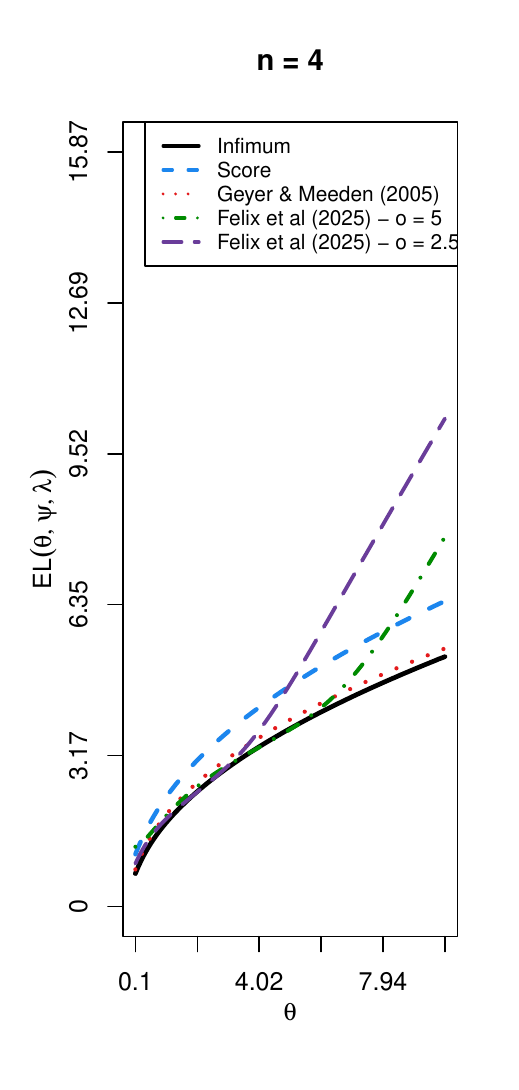}
\includegraphics[width=0.24\textwidth]{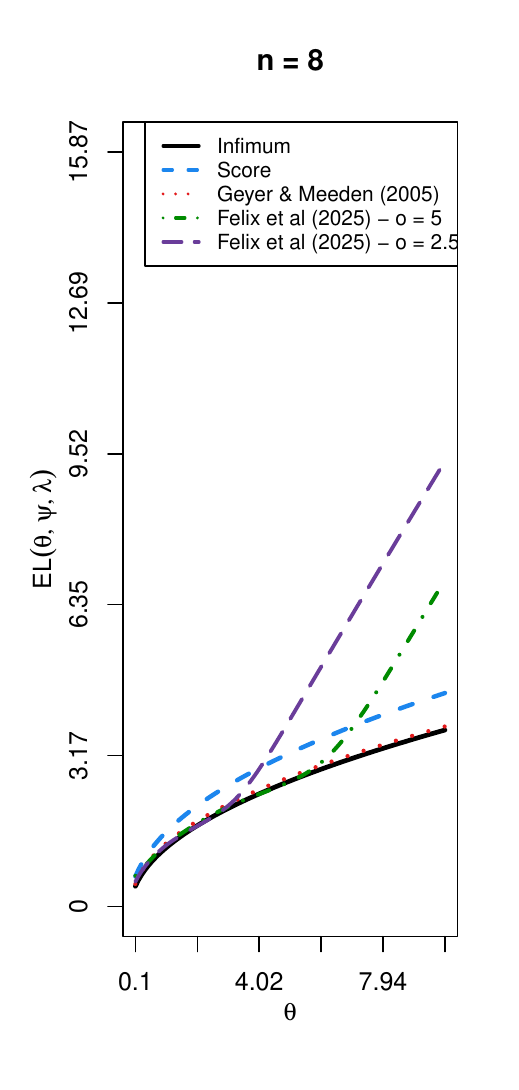}
\caption{Expected lengths for Poisson mean for the proposed method with $o = 5$ and $o = 2.5$, as well as the score and Geyer--Meeden methods. The plots were generated using the \texttt{plot.psi} function with a confidence level of $\gamma = 0.95$ and sample sizes $n = 1, 2, 4,$ and $8$.}\label{exppois_plot}
\end{figure}

We now provide examples of the membership functions for the mean of the Poisson distribution. In this case, the argument \texttt{distribution = "poisson"} is specified in the \texttt{psi} function. For the proposed method, we consider $o=5$ and $o=2.5$, respectively, and again take sample sizes $n=1,2,4,$ and $8$, as follows:

\begin{verbatim}
R> psi_poisson_o_5_1<-psi(distribution = "poisson",method = 5,n = 1)
R> psi_poisson_o_5_2<-psi(distribution = "poisson",method = 5,n = 2)
R> psi_poisson_o_5_4<-psi(distribution = "poisson",method = 5,n = 4)
R> psi_poisson_o_5_8<-psi(distribution = "poisson",method = 5,n = 8)

R> psi_poisson_o_2.5_1<-psi(distribution = "poisson",method = 2.5,n = 1)
R> psi_poisson_o_2.5_2<-psi(distribution = "poisson",method = 2.5,n = 2)
R> psi_poisson_o_2.5_4<-psi(distribution = "poisson",method = 2.5,n = 4)
R> psi_poisson_o_2.5_8<-psi(distribution = "poisson",method = 2.5,n = 8)
\end{verbatim}
Similarly to the binomial proportion case, the Geyer--Meeden and Score 
methods can be specified in the \texttt{psi} function by setting 
\texttt{method = "GM"} and \texttt{method = "Score"}, respectively:
\begin{verbatim}
R> psi_poisson_gm_1<-psi(distribution = "poisson",method = "GM",n = 1)
R> psi_poisson_gm_2<-psi(distribution = "poisson",method = "GM",n = 2)
R> psi_poisson_gm_4<-psi(distribution = "poisson",method = "GM",n = 4)
R> psi_poisson_gm_8<-psi(distribution = "poisson",method = "GM",n = 8)

R> psi_poisson_ac_1<-psi(distribution = "poisson",method = "Score",n = 1)
R> psi_poisson_ac_2<-psi(distribution = "poisson",method = "Score",n = 2)
R> psi_poisson_ac_4<-psi(distribution = "poisson",method = "Score",n = 4)
R> psi_poisson_ac_8<-psi(distribution = "poisson",method = "Score",n = 8)

\end{verbatim}
The plots of the membership functions and the corresponding expected lengths are obtained similarly to the binomial proportion, using the \texttt{plot.psi} function:

\begin{verbatim}
R> plot(psi_poisson_ac_1,psi_poisson_ac_2,psi_poisson_ac_4,psi_poisson_ac_8,
     psi_poisson_gm_1,psi_poisson_gm_2,psi_poisson_gm_4,psi_poisson_gm_8,
     psi_poisson_o_5_1,psi_poisson_o_5_2,psi_poisson_o_5_4,psi_poisson_o_5_8,
     psi_poisson_o_2.5_1,psi_poisson_o_2.5_2,psi_poisson_o_2.5_4,psi_poisson_o_2.5_8)

R> plot(psi_poisson_ac_1,psi_poisson_ac_2,psi_poisson_ac_4,psi_poisson_ac_8,
     psi_poisson_gm_1,psi_poisson_gm_2,psi_poisson_gm_4,psi_poisson_gm_8,
     psi_poisson_o_5_1,psi_poisson_o_5_2,psi_poisson_o_5_4,psi_poisson_o_5_8,
     psi_poisson_o_2.5_1,psi_poisson_o_2.5_2,psi_poisson_o_2.5_4,psi_poisson_o_2.5_8,
     type = "length")
\end{verbatim}
The resulting plots are shown in Figures \ref{mempois_plot} and 
\ref{exppois_plot}, respectively. As in 
the binomial case, the membership functions become more localized as 
$n$ increases, reflecting the additional information provided by larger 
samples. In particular, for $n=1$, the membership functions are relatively 
broad and vary smoothly with $\tau$, whereas for larger sample sizes they 
exhibit more sharply defined regions.

The location of the membership functions also changes according to the 
observed value $\omega$. As $\omega$ increases, the regions of higher 
membership tend to shift toward larger values of $\tau$, consistently 
reflecting the information provided by the observed data about the 
Poisson mean. Moreover, the effect of increasing $n$ is particularly 
evident from the progressive concentration of these regions.

A distinctive feature of the proposed method is again observed around 
the reference value $o=5$. The membership functions exhibit abrupt 
changes around $\tau=o$, especially for larger sample sizes, reflecting 
the role of $o$ in the construction of the fuzzy confidence sets. Thus, 
similarly to the binomial case, the figure illustrates how the proposed 
procedure incorporates both the observed data and the reference value 
$o$ in determining the shape and location of the membership functions.

In terms of the expected lengths, as expected from the theoretical construction, the proposed method attains the infimum at $\theta=o$. More generally, the proposed 
curves remain close to the infimum in a neighborhood of their respective 
reference values, illustrating the local optimality of the procedure. 
The choice of $o$ therefore determines the region of the parameter space 
in which the expected length is optimized.

The comparison with the alternative procedures also shows that the Score 
method produces noticeably larger expected lengths over a substantial 
part of the parameter space. The Geyer--Meeden method is closer to the 
infimum, but still exhibits larger expected lengths than the proposed 
method in relevant regions. As $\theta$ moves away from the reference 
value, however, the expected length of the proposed procedure may depart 
from the infimum, as can be observed more clearly for $o=2.5$ at larger 
values of $\theta$. This behavior illustrates the trade-off associated 
with the choice of the reference value $o$: the proposed method achieves 
optimal expected length at $\theta=o$, at the cost of potentially reduced 
efficiency farther from this value.

\subsection{Normal mean}
\begin{figure}[ht]
\includegraphics[width=0.32\textwidth]{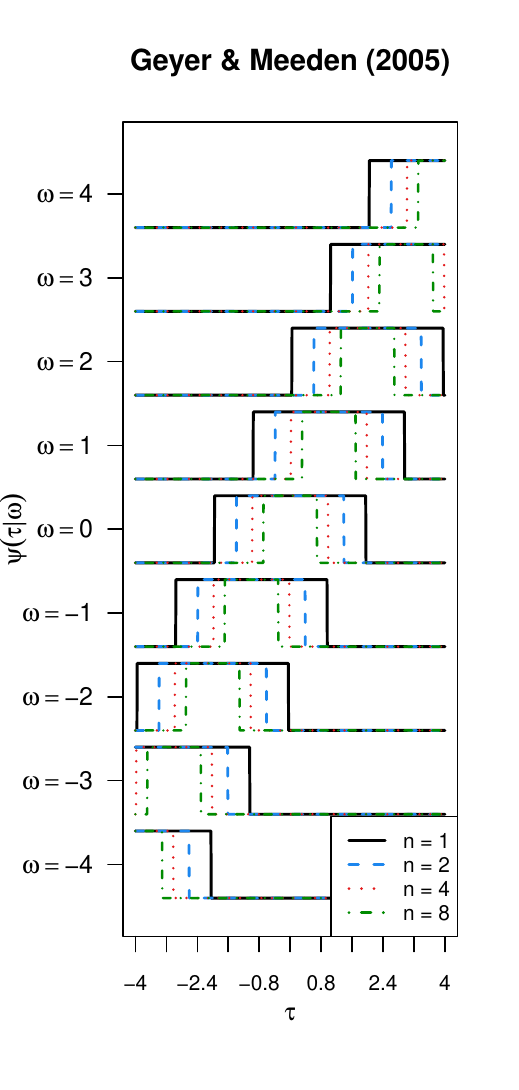}
\includegraphics[width=0.32\textwidth]{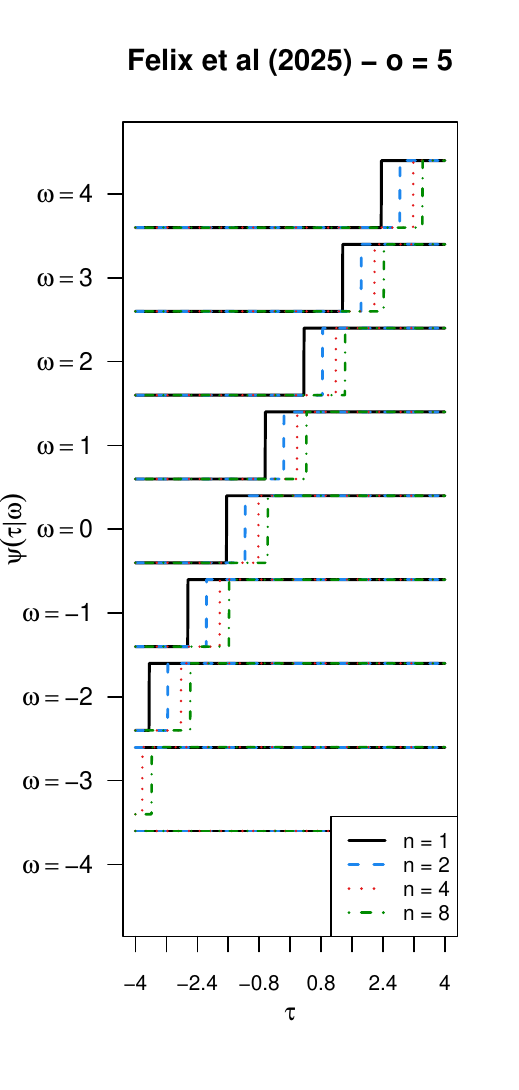}
\includegraphics[width=0.32\textwidth]{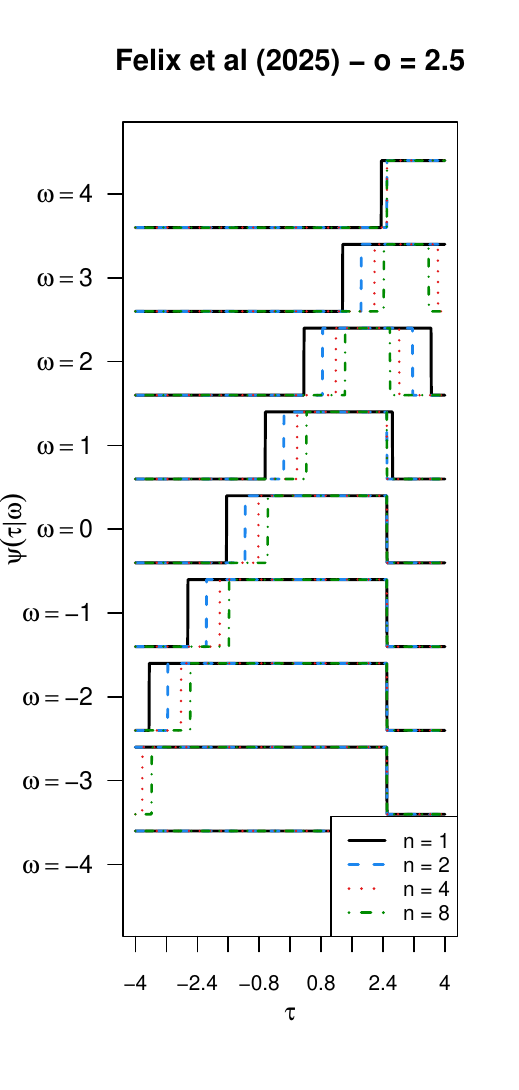}
\caption{Membership functions for the normal mean obtained using the proposed method with $o = 5$ and $o = 2.5$, as well as the Geyer--Meeden method. The plots were generated using the \texttt{plot.psi} function with a confidence level of $\gamma = 0.95$ and sample sizes $n = 1, 2, 4,$ and $8$.
}\label{membnormal_plot}
\end{figure}

\begin{figure}[h]
\includegraphics[width=0.24\textwidth]{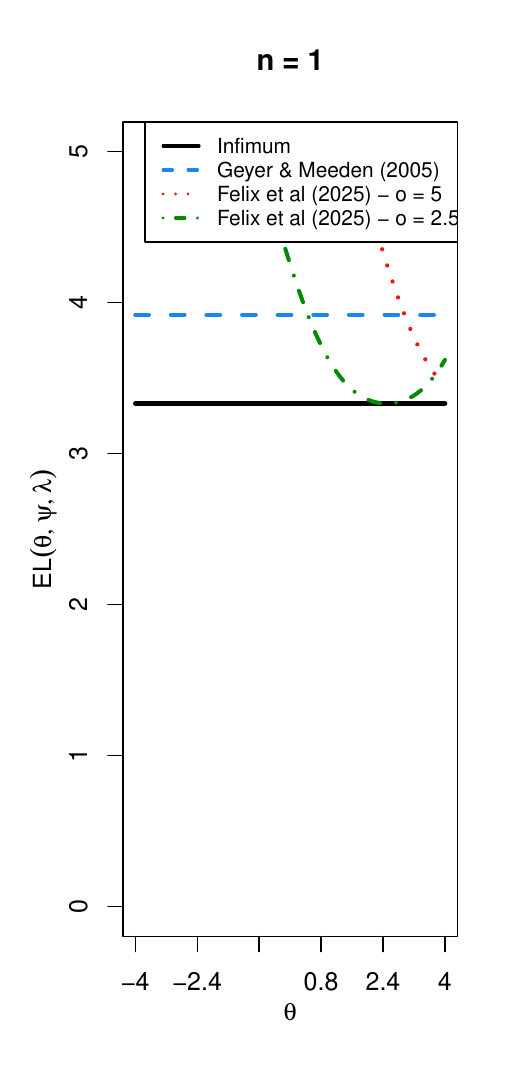}
\includegraphics[width=0.24\textwidth]{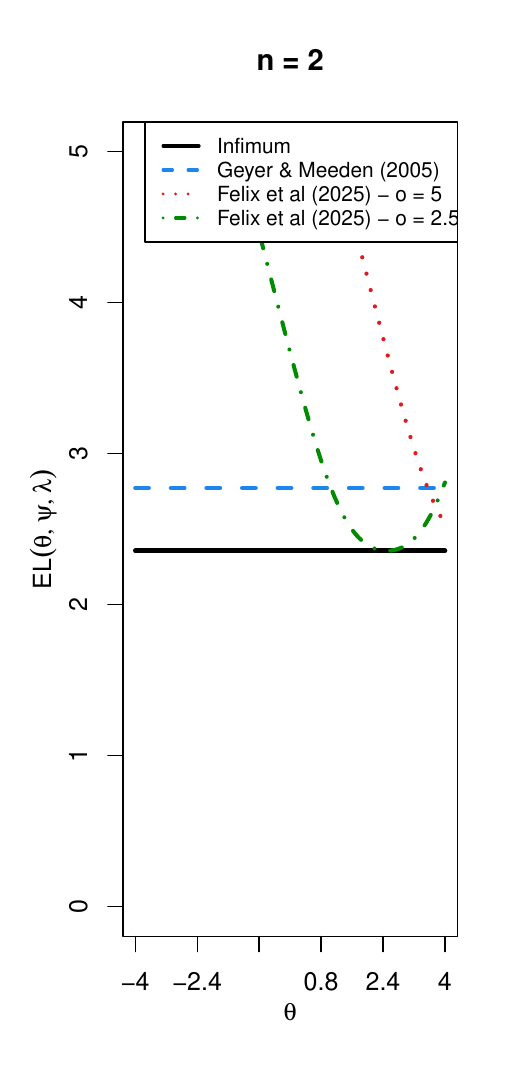}
\includegraphics[width=0.24\textwidth]{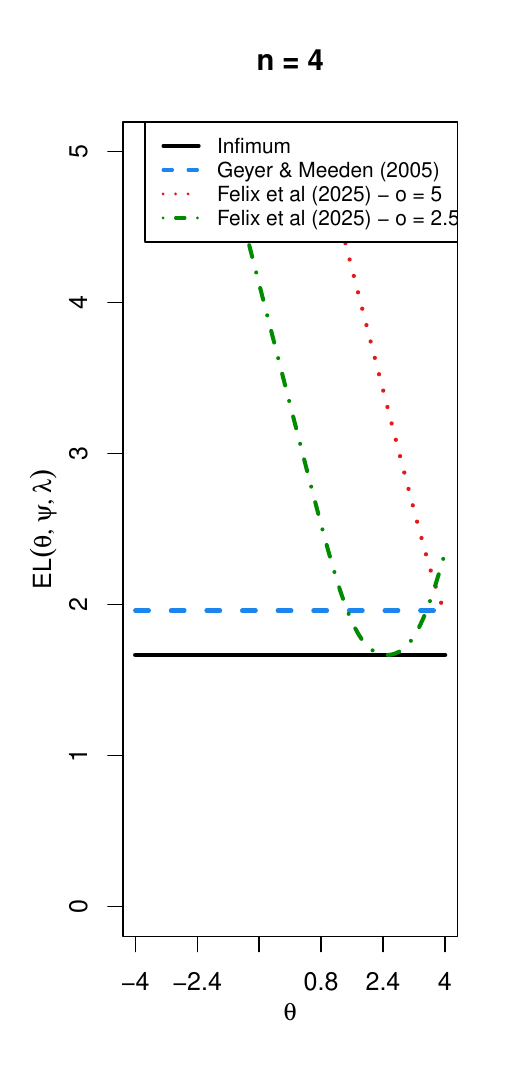}
\includegraphics[width=0.24\textwidth]{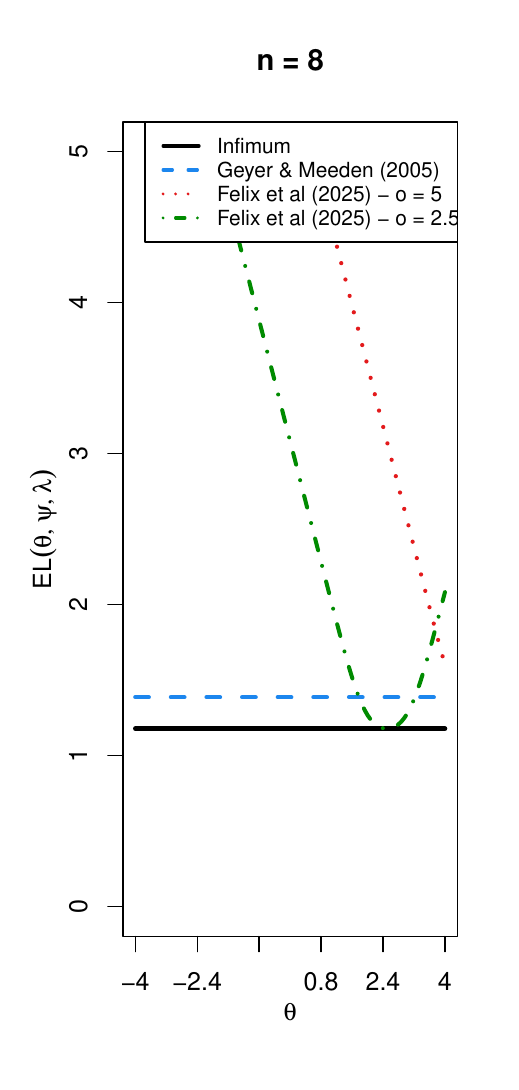}
\caption{Expected lengths for normal mean for the proposed method with $o = 5$ and $o = 2.5$, as well as the Geyer--Meeden method. The plots were generated using the \texttt{plot.psi} function with a confidence level of $\gamma = 0.95$ and sample sizes $n = 1, 2, 4,$ and $8$.} \label{expnormal_plot}
\end{figure}

Finally, for the mean of the normal distribution, we specify the argument \texttt{distribution = "normal"} in the \texttt{psi} function. For the proposed method with $o = 5$ and $2.5$ and $n = 1,2,4$ and $8$, we write:
\begin{verbatim}
R> psi_normal_o_5_1<-psi(distribution = "normal",method = 5,n = 1)
R> psi_normal_o_5_2<-psi(distribution = "normal",method = 5,n = 2)
R> psi_normal_o_5_4<-psi(distribution = "normal",method = 5,n = 4)
R> psi_normal_o_5_8<-psi(distribution = "normal",method = 5,n = 8)

R> psi_normal_o_2.5_1<-psi(distribution = "normal",method = 2.5,n = 1)
R> psi_normal_o_2.5_2<-psi(distribution = "normal",method = 2.5,n = 2)
R> psi_normal_o_2.5_4<-psi(distribution = "normal",method = 2.5,n = 4)
R> psi_normal_o_2.5_8<-psi(distribution = "normal",method = 2.5,n = 8)
\end{verbatim}
and for Geyer--Medden, we have:
\begin{verbatim}
R> psi_normal_gm_1<-psi(distribution = "normal",method = "GM",n = 1)
R> psi_normal_gm_2<-psi(distribution = "normal",method = "GM",n = 2)
R> psi_normal_gm_4<-psi(distribution = "normal",method = "GM",n = 4)
R> psi_normal_gm_8<-psi(distribution = "normal",method = "GM",n = 8)
\end{verbatim}
The plots of the membership functions and the expected lengths are obtained as:
\begin{verbatim}
R> plot(psi_normal_gm_1,psi_normal_gm_2,psi_normal_gm_4,psi_normal_gm_8,
     psi_normal_o_5_1,psi_normal_o_5_2,psi_normal_o_5_4,psi_normal_o_5_8,
     psi_normal_o_2.5_1,psi_normal_o_2.5_2,psi_normal_o_2.5_4,psi_normal_o_2.5_8)

R> plot(psi_normal_gm_1,psi_normal_gm_2,psi_normal_gm_4,psi_normal_gm_8,
     psi_normal_o_5_1,psi_normal_o_5_2,psi_normal_o_5_4,psi_normal_o_5_8,
     psi_normal_o_2.5_1,psi_normal_o_2.5_2,psi_normal_o_2.5_4,psi_normal_o_2.5_8,
     type = "length")
\end{verbatim}
which respective plots are in Figures \ref{membnormal_plot} and \ref{expnormal_plot}. In contrast to the binomial and Poisson cases, the membership
functions exhibit a more pronounced interval-like shape, with relatively
sharp transitions between regions of different membership levels.

The location of these regions changes systematically with the observed
value $\omega$. As $\omega$ increases, the corresponding membership
functions shift toward larger values of $\tau$, reflecting the information
provided by the observation about the location of the normal mean. Moreover,
for a fixed $\omega$, increasing the sample size changes the boundaries of
the regions of higher membership, illustrating the effect of the additional
sample information on the resulting fuzzy confidence set.

The influence of the reference value $o$ is also clearly visible for the proposed method.
Several membership functions exhibit a common boundary or an abrupt change
around $\tau=o$, reflecting the role of the reference value in the
construction of the proposed procedure. Overall, the figure shows that the
membership functions adapt to both the observed value and the sample size,
while retaining the characteristic behavior induced by the choice of $o$.

As expected from the theoretical construction, the expected lengths of the proposed method attains the infimum at $\theta=o$. It increases
as $\theta$ moves away from the reference value, illustrating the local
nature of the optimality property. This behavior is even more pronounced
for $o=5$. In this case, the expected length is considerably larger for
values of $\theta$ far from $o$ and progressively decreases as $\theta$
approaches the reference value.

In contrast, the Geyer--Meeden method exhibits a stable expected
length over the considered parameter space, although its curve remains
above the infimum. Therefore, the results illustrate the trade-off
inherent in the proposed construction: the choice of $o$ guarantees the
minimum expected length at $\theta=o$, but the gain around the reference
value may be accompanied by an increase in expected length for parameter
values farther from $o$. This effect is particularly evident in the
normal case considered here.

\begin{figure}[h]
\includegraphics[width=0.24\textwidth]{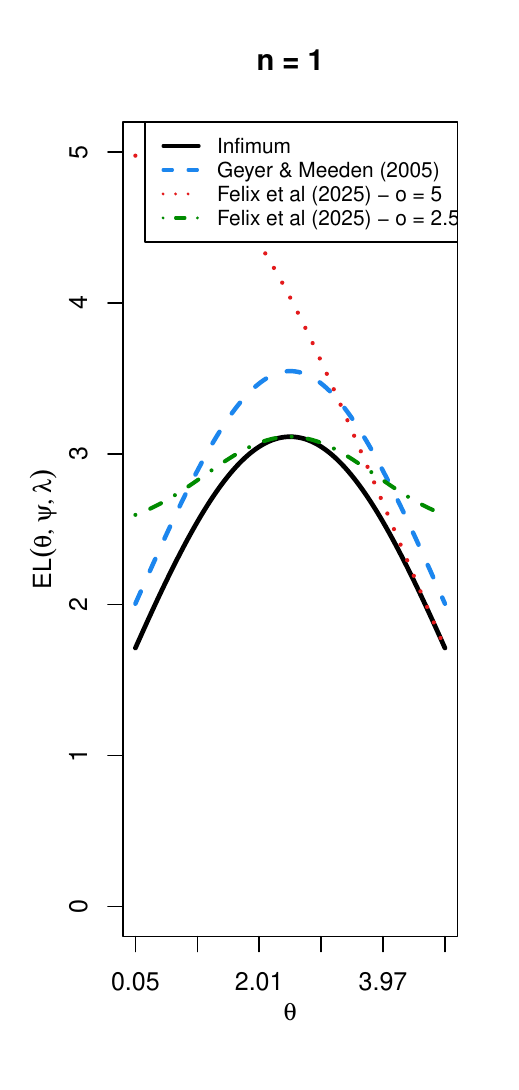}
\includegraphics[width=0.24\textwidth]{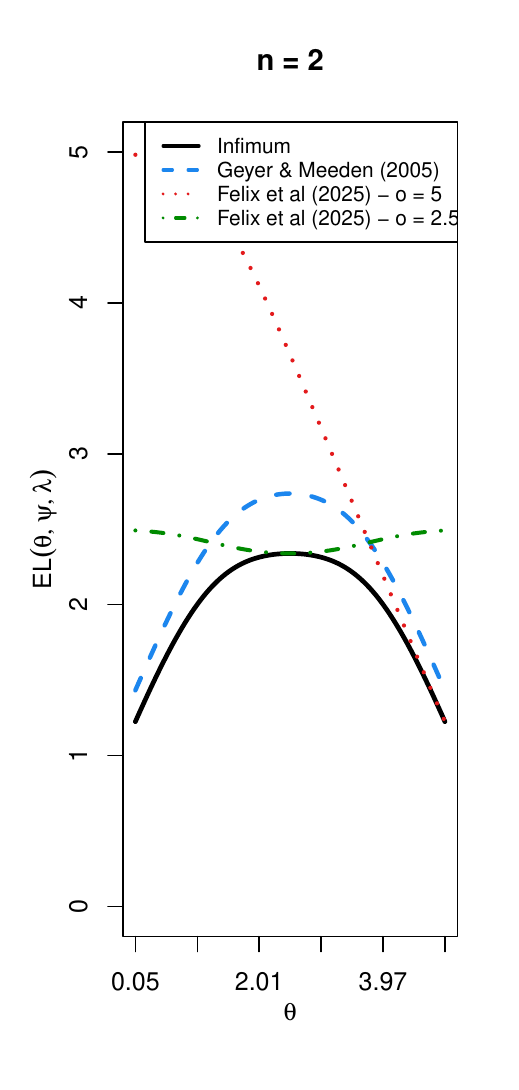}
\includegraphics[width=0.24\textwidth]{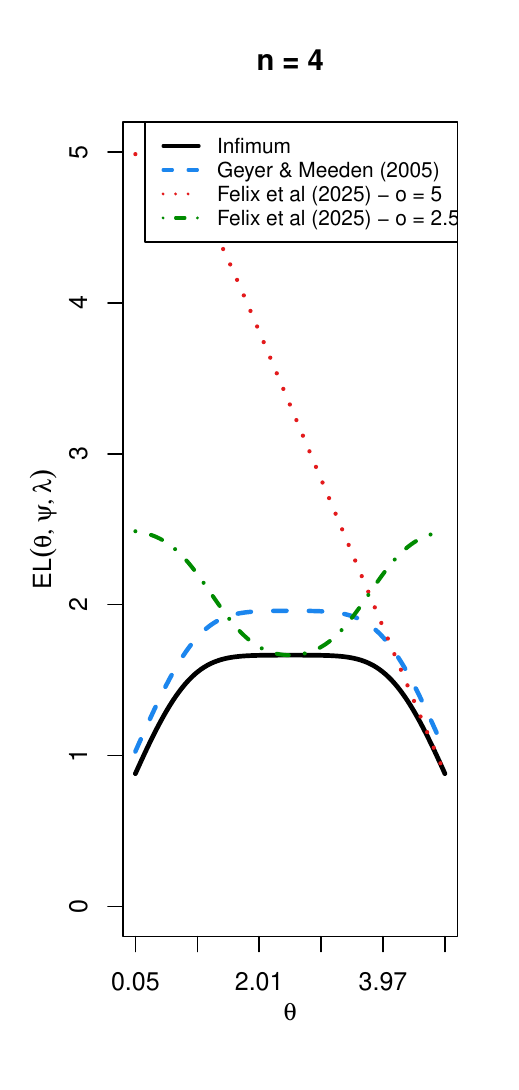}
\includegraphics[width=0.24\textwidth]{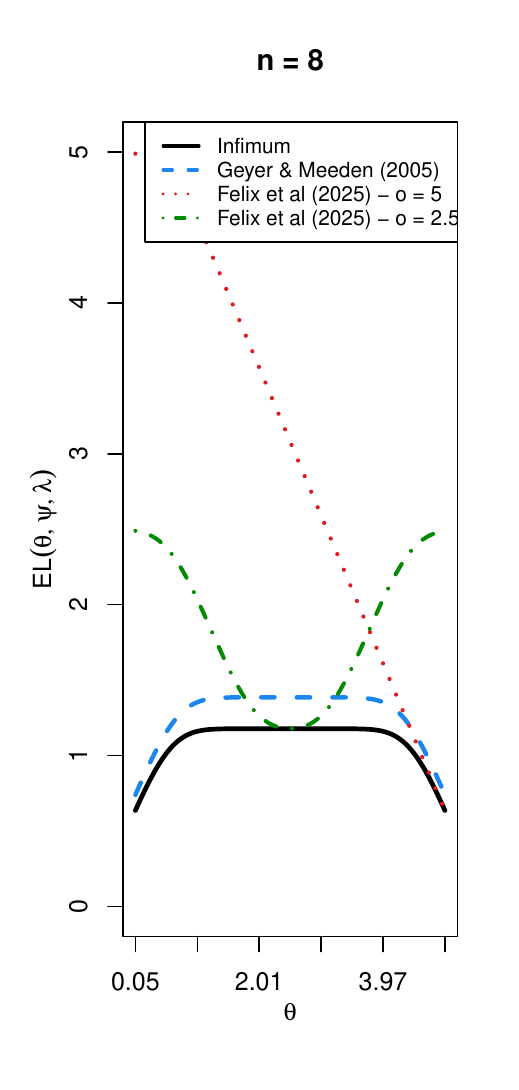}
\caption{Expected lengths for normal bounded mean for the proposed method with $o = 5$ and $o = 2.5$, as well as the Geyer--Meeden method. The plots were generated using the \texttt{plot.psi} function with a confidence level of $\gamma = 0.95$ and sample sizes $n = 1, 2, 4,$ and $8$.} \label{expbounormal_plot}
\end{figure}

The package also allows the boundedness of the normal mean to be incorporated into the analysis. In this example, we assume that $0 \leq \theta \leq 5$ by setting \texttt{tlim = c(0,5)} in the \texttt{plot.psi} function. The resulting plots of the expected lengths are shown in Figure \ref{expbounormal_plot}. As observed in the previous examples, the proposed method attains the minimum expected length at $\theta=o$, while its performance deteriorates as the parameter moves away from the reference value. Moreover, the Geyer-Meeden method exhibits a similar pattern, approaching the minimum expected length but not attaining it.
\begin{verbatim}
R> plot(psi_normal_gm_1,psi_normal_gm_2,psi_normal_gm_4,psi_normal_gm_8,
     psi_normal_o_2.5_1,psi_normal_o_2.5_2,psi_normal_o_2.5_4,psi_normal_o_2.5_8,
     psi_normal_o_5_1,psi_normal_o_5_2,psi_normal_o_5_4,psi_normal_o_5_8,
     type = "length",tlim = c(0,5),ylim = c(0,5))
\end{verbatim}

\section{Summary and discussion} \label{sec:summary}
This work presents the \texttt{FRCI} R package for constructing fuzzy and
randomized confidence intervals for the binomial proportion and the means
of the Poisson and normal distributions. The package provides two 
functions: \texttt{psi()}, which computes the membership functions and
the corresponding confidence intervals, and \texttt{plot.psi()}, which
provides graphical representations of the membership functions and their
expected lengths. The package implements the method proposed by
\cite{carlos2025}, the Geyer--Meeden method \cite{Geyer2005}, the
Agresti--Coull method \cite{Agresti1998}, and the Score method, allowing
users to construct and compare fuzzy confidence intervals under different
approaches.

Overall, the examples for the binomial, Poisson, and normal distributions
illustrate the main features of the proposed fuzzy confidence sets. The
membership functions adapt to the observed data and to the sample size,
with larger samples generally providing more sharply defined information
about the parameter. In addition, their shapes reflect the role of the
reference value $o$ in the construction of the proposed procedure.

The expected length results confirm the main optimality property of the
proposed method. In particular, for a given reference value, the
expected length attains the infimum at the reference value. Moreover, the proposed
method generally remains close to the infimum in a neighborhood of the
reference value, providing shorter expected lengths than the competing
methods in relevant regions of the parameter space.

On the other hand, this improvement is local rather than uniform over the
entire parameter space. As the parameter moves away from the reference value, the expected
length of the proposed confidence set may increase relative to the
alternative procedures. This behavior is particularly evident in the
normal case, where the choice of a reference value far from the true
parameter can lead to a substantial increase in expected length. Thus,
the choice of the reference value determines the region in which the proposed procedure
is most efficient.

These results highlight the main trade-off of the proposed approach:
it achieves the minimum expected length at a prespecified parameter value
while maintaining the required confidence level, at the cost of potentially
larger expected lengths farther from that value. Therefore, the method is
particularly attractive when prior information or the context of the
application identifies a parameter value, or a neighborhood of values,
of particular interest.

\section*{Computational details}

The results in this paper were obtained using
\texttt{R}~4.3.3 with the packages
\texttt{ump}~0.5-8,
\texttt{zipfR}~0.6-70,
\texttt{calculus}~1.1.0,
\texttt{DescTools}~0.99.60,
\texttt{graphics}~4.3.3 and
\texttt{stats}~4.3.3. \texttt{R} itself
and all packages used are available from the Comprehensive
\texttt{R} Archive Network (CRAN) at
\url{https://CRAN.R-project.org/}.


\section*{Acknowledgments}
This study was financed in part by the Coordenação de Aperfeiçoamento de Pessoal de Nível Superior – Brasil (CAPES) – Finance Code 001.  N.L.G. was financed by FAPESP grants 2019/04535-2 and 2023/13453-5, and CNPq grants 304148/2020-2 and 306496/2024-0.

\bibliographystyle{plainnat}
\bibliography{refs}


\end{document}